\documentclass[aps,pra,twocolumn,showpacs,floatfix]{revtex4}
\usepackage{bbm,amsmath,amssymb,amsbsy,graphicx}

\usepackage{txfonts}
\usepackage{type1cm}

\newcommand{\gr}[1]{\boldsymbol{#1}}
\newcommand{\sig}{\gr\sigma}
\newcommand{\ket}[1]{|#1\rangle}
\newcommand{\bra}[1]{\langle#1|}

\newcommand{\R}{\mathbbm R}
\newcommand{\C}{\mathbbm C}

\renewcommand{\det}{{\rm Det}\,}
\newcommand{\eq}[1]{Eq.~(\ref{#1})}
\newcommand{\ineq}[1]{Ineq.~(\ref{#1})}

\begin{document}
\title{Coexistence of unlimited bipartite and genuine multipartite entanglement:
\\ Promiscuous quantum correlations arising from discrete to continuous
variable systems}
\date{May 12, 2006}
\author{Gerardo Adesso$^{1,2}$, Marie Ericsson$^2$, and Fabrizio Illuminati$^1$}

\affiliation{$^1$Dipartimento di Fisica ``E. R. Caianiello'',
Universit\`a degli Studi di Salerno, INFN Sezione di Napoli-Gruppo
Collegato di Salerno, Via S. Allende, 84081 Baronissi (SA), Italy
\\ $^2$Centre for Quantum Computation, DAMTP, Centre for
Mathematical Sciences, University of Cambridge, Wilberforce Road,
Cambridge CB3 0WA, United Kingdom}

\begin{abstract}
Quantum mechanics imposes `monogamy' constraints on the sharing of
entanglement. We show that, despite these limitations, entanglement
can be fully `promiscuous', i.e.~simultaneously present in unlimited
two-body and many-body forms in states living in an
infinite-dimensional Hilbert space. Monogamy just bounds the
divergence rate of the various entanglement contributions. This is
demonstrated in simple families of $N$-mode ($N \ge 4$) Gaussian
states of light fields or atomic ensembles, which therefore enable
infinitely more freedom in the distribution of information, as
opposed to systems of individual qubits. Such a finding is of
importance for the quantification, understanding and potential
exploitation of shared quantum correlations in continuous variable
systems. We discuss how promiscuity gradually arises when
considering simple families of discrete variable states, with
increasing Hilbert space dimension towards the continuous variable
limit. Such models are somehow analogous to Gaussian states with
asymptotically diverging, but finite squeezing. In this respect, we
find that non-Gaussian states (which in general are more entangled
than Gaussian states), exhibit also the interesting feature that
their entanglement is more shareable: in the non-Gaussian
multipartite arena, unlimited promiscuity can be already achieved
among three entangled parties, while this is impossible for
Gaussian, even infinitely squeezed states.
\end{abstract}

\maketitle

\section{Introduction}

Entanglement is a core concept in quantum mechanics \cite{schr}, and
a key resource for quantum communication and information processing
\cite{book}. Unlike classical correlations, entanglement cannot be
freely distributed \cite{pisa}. In a multipartite compound system of
two-level quantum objects (qubits), if two subsystems are maximally
entangled, they cannot share any residual form of quantum
correlations with the other remaining parties \cite{ckw,osborne}.
Analogous {\em monogamy} relations have been recently established
for entanglement between canonical conjugate variables of continuous
variable (CV) systems \cite{contangle,hiroshima}, like harmonic
oscillators, light modes and atomic ensembles, endowed with an
infinite-dimensional Hilbert space. In the general case of a state
distributed among $N$ parties (each owning a single qubit, or a
single mode, respectively), the monogamy constraint on bipartite
entanglement takes the form of the Coffman-Kundu-Wootters inequality
\cite{ckw}
\begin{equation}
\label{ckwine} E_{S_i \vert (S_1 \ldots S_{i-1} S_{i+1} \ldots S_N
)} \ge \sum\limits_{j\ne i}^N {E_{S_i \vert S_j } }
\end{equation}
where the global system is multipartitioned in subsystems $S_k$
($k=1,{\ldots},N$), each owned by a respective party, and $E$ is a
proper measure of bipartite entanglement. The left-hand side of
inequality (\ref{ckwine}) quantifies the bipartite entanglement
between a probe subsystem $S_i $ and the remaining subsystems taken
as a whole. The right-hand side quantifies the total bipartite
entanglement between $S_i$ and each one of the other subsystems
$S_{j\ne i}$ in the respective reduced states. The non-negative
difference between these two entanglements, minimized over all
choices of the probe subsystem, is referred to as the
\textit{residual multipartite entanglement}. It quantifies the
purely quantum correlations that are not encoded in pairwise form,
so it includes all manifestations of genuine $K$-partite
entanglement, involving $K$ subsystems at a time, with $2<K\le N$.
In the simplest nontrivial instance of $N=3$, the residual
entanglement has the (physical and mathematical) meaning of the
genuine tripartite entanglement shared by the three subsystems
\cite{ckw,wstates,contangle}. In the general case $N>3$, a natural
way to discriminate between all those entanglement contributions has
been very recently advanced in Ref.~\cite{strongmono}, and proven
successful when addressing entanglement shared by harmonic systems
under complete permutation invariance. The study of entanglement
sharing and monogamy constraints thus offers a natural framework to
interpret and quantify entanglement in multipartite quantum systems
\cite{pisa}.

From an operational perspective, qubits are the main logical units
for standard realizations of quantum information (QI) protocols
\cite{book}. Also CV Gaussian entangled resources have  been proven
useful for all known implementations of QI processing
\cite{brareview}, including quantum computation \cite{menicucci},
sometimes outperforming more traditional qubit-based approaches as
in the case of teleportation \cite{furuscience}. It is therefore
important to understand if special features of entanglement appear
in states of infinite Hilbert spaces, which are unparalleled in the
corresponding states of qubits. Such findings may lead to new ways
of manipulating QI in the CV setting \cite{slocc}, and may in
general contribute to a deeper and more complete understanding of
entanglement in complex systems.

 In this paper, we address this motivation
on a clear-cut physical ground, aiming in particular to show whether
the unboundedness of the mean energy characterizing CV states
enables a qualitatively richer structure for distributed quantum
correlations. We prove that multimode Gaussian states exist, that
can possess simultaneously arbitrarily large pairwise bipartite
entanglement between some pairs of modes and arbitrarily large
genuine multipartite entanglement among all modes. In particular, we
focus on a four-mode family of Gaussian states which are producible
with standard optical means, the achievable amount of entanglement
being technologically limited only by the attainable degree of
squeezing. These states asymptotically reach the form of two
perfectly entangled Einstein-Podolsky-Rosen (EPR) pairs that can
moreover be arbitrarily intercorrelated quantumly, as well as share
an  asymptotically diverging genuine four-partite entanglement among
all modes. We show that this {\em promiscuity} is fully compatible
with the monogamy inequality~(\ref{ckwine}) for CV entanglement as
established in Refs.~\cite{contangle,hiroshima}, and even with the
stronger monogamy constraints on distributed bipartite and
multipartite quantum correlations recently put forward in
Ref.~\cite{strongmono}. This original feature of entanglement sheds
new light on the actual extent to which the basic laws of quantum
mechanics curtail the distribution of information; on a more
applicative ground, it may serve as a prelude to implementations of
QI processing in a infinite-dimensional scenario that {\em cannot}
be achieved with single qubit resources. While promiscuity might
seem an  unexpected property of quantum correlations, we will show
that it indeed arises naturally and gradually with increasing
dimension of the Hilbert space associated to the
entanglement-sharing parties. In the limit of continuous variable
systems, non-Gaussian states will be shown to be in general sensibly
more powerful than Gaussian states from the point of view of
(promiscuous) entanglement sharing.


\section{Entanglement sharing among Gaussian modes}

Entanglement in CV systems is encoded in the form of EPR
correlations \cite{epr}. Let us consider the motional degrees of
freedom of two particles, or the quadratures of a two-mode radiation
field, where mode $k=i,j$ is described by the ladder operators $\hat
{a}_k ,\,\hat {a}_k ^\dag $ satisfying the bosonic commutation
relation $[\hat {a}_k ,\,\hat {a}_k ^\dag ]=1$. An arbitrarily
increasing degree of entanglement
 \cite{noteunlim} (and of mean energy) can be encoded in a two-mode
squeezed state $\ket{\psi^{sq}}_{i,j}=U_{i,j}(r)
\left(\ket{0}_i\!\otimes\ket{0}_j\right)$ with increasing squeezing
factor $r \in \R$, where  the (phase-free) two-mode squeezing
operator is \cite{qopt}  $U_{i,j}(r) = \exp \left[\frac{r}{2} (\hat
{a}_i^\dag \hat {a}_j^\dag -\hat {a}_i \hat {a}_j ) \right]$, and
$\left| 0 \right\rangle _k$ denotes the vacuum state in the Fock
space of mode $k$. In the limit of infinite squeezing ($r\to \infty
)$, the state approaches the ideal (unnormalizable, infinite-energy
and unphysical) EPR state \cite{epr}, simultaneous eigenstate of
total momentum and relative position of the two subsystems, which
thus share infinite entanglement. The entanglement of squeezed
states and, more generally,  {\em Gaussian states}, has been
intensively studied in recent times \cite{adebook}. Gaussian states
of $N$ modes are completely described in phase space by the
$2N\times 2N$ real symmetric covariance matrix (CM) $\sig$ of the
second moments $\textstyle{1 \over 2}\left\langle {\hat {X}_i \hat
{X}_j +\hat {X}_j \hat {X}_i } \right\rangle $ of the canonical
bosonic operators $\hat {X}_k \equiv \hat {q}_k =\hat {a}_k +\,\hat
{a}_k ^\dag ,\,\hat {X}_{N+k} \equiv \hat {p}_k =\hat {a}_k -\,\hat
{a}_k ^\dag (k=1\ldots N)$. In this representation the unitary
two-mode squeezing operator amounts to a symplectic matrix $S_{i,j}
(r)={{c_r\ s_r}\choose{s_r\ c_r}}\oplus{{\ c_r\ -s_r}\choose{-s_r\ \
c_r}}$ (where $c_r =\cosh r,\,s_r =\sinh r$, which acts by
congruence on the CM, $\sig \mapsto S \sig S^T$.

Entanglement sharing has been addressed for Gaussian states in
Refs.~\cite{contangle,hiroshima,strongmono}, by adapting and
extending the original Coffman-Kundu-Wootters analysis \cite{ckw}
(see also \cite{relaxandshare} where a different analysis is
performed, which nonetheless has implications for the sharing of
correlations in multimode Gaussian states). In the most basic
multipartite CV setting, namely that of three-mode Gaussian states,
a partial ``promiscuity'' of entanglement can be achieved.
Permutation-invariant states exist which are the Gaussian
simultaneous analogues of Greenberger-Horne-Zeilinger (GHZ) and $W$
states of qubits \cite{ghzs,wstates}. They possess unlimited
tripartite entanglement (with increasing squeezing) and nonzero,
accordingly increasing bipartite entanglement which nevertheless
stays finite even for infinite squeezing \cite{contangle}. We will
now show that in CV systems with more than three modes, entanglement
can be distributed in an {\em infinitely} promiscuous way. To
illustrate the existence of such phenomenon, we consider the
simplest nontrivial instance of a family of four-mode Gaussian
states, endowed with a partial symmetry under mode exchange.

\section{Four-mode states: structural and entanglement properties}

\begin{figure}[t!]
\includegraphics[width=8cm]{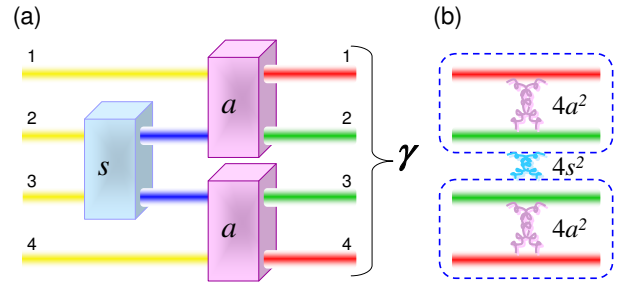} \caption{(Color online)
(a) Preparation of the four-mode Gaussian states $\gr\gamma$ of
\eq{s4} via two-mode squeezers (boxes) applied on vacuum modes
(yellow beams). (b) Structure of bipartite entanglement.}
\label{figprep}
\end{figure}

We start with an uncorrelated state of four modes, each one
initially in the vacuum of the respective Fock space, whose
corresponding CM is the identity. We apply a two-mode squeezing
transformation with squeezing $s$ to modes 2 and 3, then two further
two-mode squeezing transformations (redistributing the initial
pairwise entanglement among all modes) with squeezing $a$ to the
pairs of modes 1,2 and 3,4 [see Fig.~\ref{figprep}(a)]. For any
value of the parameters $s$ and $a$, the output is a pure four-mode
Gaussian state  with CM $\gr\gamma$,
\begin{equation}\label{s4}
\gr\gamma =S_{3,4}(a)S_{1,2}(a)S_{2,3}(s)S_{2,3}^T (s)S_{1,2}^T
(a)S_{3,4}^T (a)
\end{equation}
A state of this form is  invariant under the double exchange of
modes $1\leftrightarrow 4$ and $2\leftrightarrow 3$, as $S_{i,j}
=S_{j,i} $ and operations on disjoint pairs of modes commute.

In a pure four-mode Gaussian state and in its reductions, bipartite
entanglement is equivalent to  negativity of the partially
transposed CM, obtained by reversing time in the subspace of any
chosen single subsystem \cite{simon00,werwolf}. This inseparability
criterion is readily verified for the family of states in \eq{s4}
yielding that, for all nonzero values of the squeezings $s$ and $a$,
$\gr\gamma$ is entangled with respect to any global bipartition of
the modes. The state is thus said to be {\em fully inseparable}
\cite{book}, i.e. it contains genuine four-partite entanglement.
Following previous studies on CV entanglement sharing we use the
``contangle'' $\tau$ \cite{contangle} to quantify bipartite
entanglement, an entanglement monotone under Gaussian local
operations and classical communication. It is defined for pure
states as the squared logarithmic negativity \cite{vidwer} (which
quantifies how much the partially transposed state fails to be
positive) and extended to mixed states via Gaussian convex roof
\cite{geof,ordering}, i.e. as the minimum of the average pure-state
entanglement over all decompositions of the mixed state in ensembles
of pure Gaussian states. If $\sig_{i\vert j} $ is the CM of a
(generally mixed) bipartite Gaussian state where subsystem $i$
comprises one mode only, then the contangle $\tau $ can be computed
as
\begin{equation}
\label{tau} \tau (\sig_{i\vert j} )\equiv \tau (\sig_{i\vert
j}^{opt} )=g[m_{i\vert j}^2 ],\;\;\;g[x]={\rm arcsinh}^2[\sqrt
{x-1}].
\end{equation}
Here $\sig_{i\vert j}^{opt} $ corresponds to a pure Gaussian state,
and $m_{i\vert j} \equiv m(\sig _{i\vert j}^{opt} )=\sqrt {\det
\sig_i^{opt} } =\sqrt {\det \sig_j^{opt}}$, with $\sig_{i(j)}^{opt}$
being the reduced CM of subsystem $i (j)$ obtained by tracing over
the degrees of freedom of subsystem $j$ ($i)$. The CM $\sig_{i\vert
j}^{opt} $ denotes the pure bipartite Gaussian state which minimizes
$m(\sig_{i\vert j}^p )$ among all pure-state CMs $\sig_{i\vert j}^p
$ such that $\sig_{i\vert j}^p \le \sig_{i\vert j}$. If
$\sig_{i\vert j}$ is a pure state, then $\sig_{i\vert j}^{opt}
=\sig_{i\vert j}$, while for a mixed Gaussian state \eq{tau} is
mathematically equivalent to constructing the Gaussian convex roof.
For a separable state $m(\sig_{i\vert j}^{opt})=1$. The contangle
$\tau$ is completely equivalent to the Gaussian entanglement of
formation \cite{geof}, which quantifies the cost of creating a given
mixed Gaussian state out of an ensemble of pure, entangled Gaussian
states.

\subsection{Structure of bipartite entanglement}

In the four-mode state with CM $\gr\gamma$, we can compute the
bipartite contangle in closed form \cite{ordering} for all pairwise
reduced (mixed) states of two modes $i$ and $j$, described by a CM
$\gr\gamma_{i\vert j} $. By applying the partial transpose criterion
\cite{simon00}, we find that the two-mode states indexed by the
partitions $1\vert3$, $2\vert4$, and $1\vert4$, are separable. For
the remaining two-mode states we find $m_{1\vert 2} =m_{3\vert 4}
=\cosh 2a$, while $m_{2\vert 3} $ is equal to $\frac{-1+2\cosh
^2(2a)\cosh ^2s+3\cosh (2s)-4\sinh ^2a\sinh (2s)}{4[\cosh
^2a+e^{2s}\sinh ^2a]}$ if $a<{\rm arcsinh}[\sqrt{\tanh s}]$, and to
$1$ (implying separability) otherwise. Accordingly, one can compute
the pure-state entanglements between one probe mode and the
remaining three modes. One finds $m_{1\vert (234)} =m_{4\vert (123)}
=\cosh ^2a+\cosh (2s)\sinh ^2a$ and $m_{2\vert (134)} =m_{3\vert
(124)} =\sinh ^2a+\cosh (2s)\cosh ^2a$.

Concerning the structure of bipartite entanglement, the contangle in
the mixed two-mode states $\gr\gamma_{1\vert 2} $ and
$\gr\gamma_{3\vert 4} $ is $4a^2$, unrespective of the value of $s$.
This quantity is exactly equal to the degree of entanglement in a
pure two-mode squeezed state $S_{i,j} (a)S_{i,j}^T (a)$ of modes $i$
and $j$ generated with the same squeezing $a$. In fact, the two-mode
mixed state $\gamma_{1\vert 2} $ (and, equivalently, $\gamma_{3\vert
4}$) serves as a proper resource for CV teleportation
\cite{furuscience}, realizing a perfect transfer (unit fidelity
\cite{notefidel}) in the limit of infinite squeezing $a$. The
four-mode state   $\gr\gamma$ reproduces thus (in the regime of very
high $a$) the entanglement content of two EPR-like pairs (1,2 and
3,4). Remarkably, there is an additional, independent entanglement
\textit{between} the two pairs, given by $\tau (\gamma _{(12)\vert
(34)} )=4s^2$ [the original entanglement in the two-mode squeezed
state $S_{2,3} (s)S_{2,3}^T (s)$ after the first construction step,
see Fig.~\ref{figprep}(a)], which can be itself increased
arbitrarily with increasing $s$. This peculiar distribution of
bipartite entanglement  [see Fig.~\ref{figprep}(b)] is a first
interesting signature of an unmatched freedom of entanglement
sharing in multimode Gaussian states as opposed for instance to
states of the same number of qubits, where a similar situation is
\textit{impossible}. Specifically, if in a pure state of four qubits
the first two approach unit entanglement and the same holds for the
last two, the only compatible global state reduces necessarily to a
product state of the two singlets: no interpair entanglement is
allowed by the  monogamy constraint \cite{ckw,osborne}.

\subsection{Residual entanglement}

We can now move to a closer analysis of entanglement distribution
and genuine multipartite quantum correlations. A first step is to
verify whether the monogamy inequality~(\ref{ckwine}) is satisfied
on the four-mode state $\gr\gamma$ distributed among the four
parties (each one owning a single mode) \cite{notemonocheck}. In
fact, the problem reduces to proving that $\min \{g[m_{1\vert
(234)}^2 ]-g[m_{1\vert 2}^2 ],\,g[m_{2\vert (134)}^2 ]-g[m_{1\vert
2}^2 ]-g[m_{2\vert 3}^2 ]\}$ is nonnegative. The first quantity
always achieves the minimum yielding
\begin{eqnarray}
\label{taures} \tau^{res}(\gr\gamma ) &\!\!\equiv\!\!& \tau
(\gr\gamma _{1\vert (234)} )-\tau (\gr\gamma _{1\vert 2} )
\\ &\!\!=\!\!&{\rm arcsinh}^2\!\left\{ {\sqrt {[\cosh
^2a+\cosh (2s)\sinh ^2a]^2-1} } \right\}-4a^2. \nonumber
\end{eqnarray}
Since $\cosh (2s)>1$ for $s>0$, it follows that $\tau^{res}>0$. The
entanglement in the global Gaussian state is therefore correctly
distributed according to the monogamy law, in such a way that the
residual contangle $\tau^{res}$ quantifies the multipartite
entanglement not stored in couplewise form. Those quantum
correlations can be either tripartite involving three of the four
modes, and/or genuinely four-partite among all of them. Concerning
the tripartite entanglement, we first observe that in the
tripartitions 1$\vert $2$\vert $4 and 1$\vert $3$\vert $4 the
tripartite entanglement is zero, as mode 4 is not entangled with the
block of modes 1,2, and mode 1 is not entangled with the block of
modes 3,4

\begin{figure}[t!]
\includegraphics[width=8.5cm]{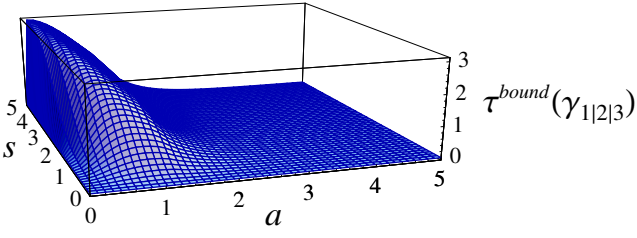} \caption{(Color online)
Upper bound $\tau ^{bound}(\gr\gamma _{1\vert 2\vert 3} )$,
\eq{taubnd}, on the tripartite entanglement between modes 1, 2 and 3
(and equivalently 2, 3, and 4) of the four-mode Gaussian state
defined by \eq{s4}, plotted as a function of the squeezing
parameters $s$ and $a$.
Focusing on the tripartition 1$\vert $2$\vert $3, the bipartite
contangle $\tau (\gr\gamma _{i\vert (jk)} )$ (with $i,j,k$ a
permutation of 1,2,3) is bounded from above by the corresponding
bipartite contangle $\tau (\gr\sigma _{i\vert (jk)}^p )$ in any
pure, three-mode Gaussian state with CM $\gr\sigma _{i\vert (jk)}^p
\le \gr\gamma _{i\vert (jk)}$. The state $\gr\sigma ^p=S_{1,2}
(a)S_{2,3} (t)S_{2,3}^T (t)S_{1,2}^T (a)$ of the three modes 1, 2
and 3, with $t=\textstyle{1 \over 2}{\rm arccosh}[\textstyle{{1+{\rm
sech}^2a\tanh ^2s} \over {1-{\rm sech}^2a\tanh ^2s}}]$, satisfies
this condition and from \eq{tau} we have $\tau (\gamma _{i\vert
(jk)} )\le g[(m_{i\vert (jk)}^{bound} )^2]$. This leads to
\eq{taubnd}, where the quantity $g[(m_{2\vert (13)}^{bound}
)^2]-g[m_{1\vert 2}^2 ]-g[m_{2\vert 3}^2 ]$ with $m_{2\vert
(13)}^{bound} =\sinh ^2a+m_{3\vert (12)}^{bound} \cosh ^2a$ is not
included in the minimization, being always larger than the other
terms.
}\label{figtrip}
\end{figure}

The only tripartite entanglement present, if any, is equal in
content (due to the symmetry of the state $\gr\gamma$) for the
tripartitions 1$\vert $2$\vert $3 and 2$\vert $3$\vert $4, and can
be quantified by the residual contangle (a Gaussian entanglement
monotone \cite{contangle}) determined by the corresponding
three-mode monogamy inequality~(\ref{ckwine}). One can show (see
caption of Fig.~\ref{figtrip}) that such genuine tripartite
entanglement $\tau^{tri}(\gr\gamma _{1\vert 2\vert 3})$ is bounded
from above by the quantity
\begin{equation}
\label{taubnd} \tau ^{bound}(\gr\gamma _{1\vert 2\vert 3} )\equiv
\min \{g[(m_{1\vert (23)}^{bound} )^2]-g[m_{1\vert 2}^2
],\,g[(m_{3\vert (12)}^{bound} )^2]-g[m_{2\vert 3}^2 ]\},\!
\end{equation}
with $m_{3\vert (12)}^{bound} =\textstyle{{1+{\rm sech}^2a\tanh ^2s}
\over {1-{\rm sech}^2a\tanh ^2s}}$, $m_{1\vert (23)}^{bound} =\cosh
^2a+m_{3\vert (12)}^{bound} \sinh ^2a$. The upper bound $\tau
^{bound}(\gr\gamma _{1\vert 2\vert 3} )$ is always nonnegative (as a
consequence of monogamy \cite{contangle}), is decreasing with
increasing squeezing $a$, and vanishes in the limit $a\to \infty $,
as shown in Fig.~(\ref{figtrip}). Therefore, in the regime of
increasingly high $a$, eventually approaching infinity, any form of
tripartite entanglement among any three modes in the state
$\gr\gamma $ is negligible (exactly vanishing in the limit). As a
crucial consequence, in that regime the residual entanglement $\tau
^{res}(\gr\gamma )$ determined by \eq{taures} is all stored in
four-mode quantum correlations and quantifies the \textit{genuine}
four-partite entanglement.

We finally observe that $\tau ^{res}(\gr\gamma)$ is an increasing
function of $a$ for any value of $s$ (see Fig.~\ref{figres}), and it
\textit{diverges} in the limit $a\to \infty $. This proves that the
class of pure four-mode Gaussian states with CM $\gr\gamma$ given by
\eq{s4} exhibits genuine four-partite entanglement which grows
unboundedly with increasing squeezing $a$ and, simultaneously,
possesses pairwise bipartite entanglement in the mixed two-mode
reduced states of modes 1,2 and 3,4, that increases unboundedly as
well with with increasing $a$ \cite{noteunlim}. Moreover, as
previously shown, the two pairs can themselves be arbitrarily
entangled with each other with increasing squeezing $s$. Notice that
usual monogamy inequalities do not typically constrain ``hybrid''
entanglement distributions such as $2 \times 2$ versus $1 \times 1$
or $1 \times 3$, even though it is reasonable to expect that some
limitations exist also when entanglement is shared over such
partitions.

\subsection{Strong monogamy and genuine four-partite entanglement}

\begin{figure}[t!]
\includegraphics[width=8.5cm]{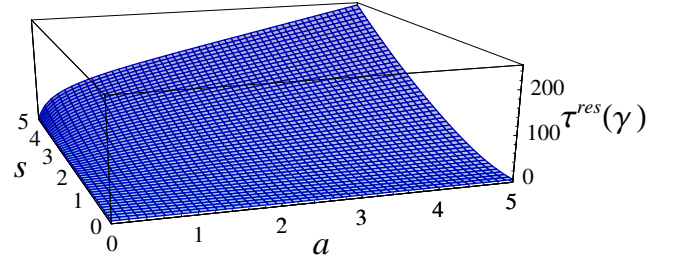} \caption{(Color online)
Residual multipartite entanglement $\tau ^{res}(\gr\gamma )$
[\eq{taures}], which in the regime of large squeezing $a$ is
completely distributed in the form of genuine four-partite quantum
correlations.
} \label{figres}
\end{figure}

It would be desirable to have a clear distinction between tripartite
and four-partite quantum correlations also in the range in which the
former are not negligible, i.e., for small $a$. A simple idea comes
from Ref.~\cite{strongmono}, which adapted to our setting states the
following. The minimum residual entanglement \eq{taures} between one
probe mode and the remaining three, is equal to the total three-mode
entanglement involving the probe mode and any two of the others,
plus the genuine four-partite entanglement shared among all modes.
Such a construction, which provides an additional decomposition for
the difference between LHS and RHS of \ineq{ckwine}, leads in
general to postulate that a {\em stronger} monogamy constraint acts
on entanglement shared by $N$ parties, which imposes a trade-off on
both the bipartite and all forms of genuine multipartite
entanglement on the same ground. Existence of such a constraint,
which directly generalizes the Coffman-Kundu-Wootters ($N=3$) case
\cite{ckw}, has been indeed proven for arbitrary  Gaussian states
endowed with permutation invariance (i.e., for the generalized
$N$-mode Gaussian GHZ/$W$ states and their mixed reductions), and a
{\em bona fide} quantification of the genuine $N$-party entanglement
(computed again by means of the contangle) has been obtained on such
states \cite{strongmono}.

We will now show that strong monogamy also holds in the partially
symmetric four-mode Gaussian states of \eq{s4}, which is an
important indication of the generality of the approach presented in
\cite{strongmono}. Indeed, strong monogamy immediately follows from
Eqs.~(\ref{taures},\ref{taubnd}), as $\tau^{res}(\gr\gamma )-\tau
^{tri}(\gr\gamma _{1\vert 2\vert 3} ) \ge \tau^{res}(\gr\gamma ) \ge
\tau ^{bound}(\gr\gamma _{1\vert 2\vert 3} ) \equiv \tau
^{4bound}(\gr\gamma _{1\vert 2\vert 3 \vert 4} ) \ge 0$. The
quantity $\tau ^{4bound}(\gr\gamma _{1\vert 2\vert 3 \vert 4} )$ is
a {\em lower} bound to the genuine four-partite entanglement shared
by the four modes in the state $\gr\gamma$, for any $a$  and $s$.
Its functional dependence on these squeezing parameters is very
similar to that of $\tau^{res}(\gr\gamma )$ (one needs only to
subtract the surface in Fig.~\ref{figtrip}, which extends over a
very narrow scale, from the surface in Fig.~\ref{figres}).
$\tau^{res}(\gr\gamma )$ represents actually an upper bound for the
four-partite entanglement, which gets asymptotically tight for $a
\gg 1$: in this limit, upper and lower bounds coincide and as
already remarked, the residual entanglement {\em is} all in
four-partite form. This brief analysis is very relevant on one hand
for the purposes of Ref.~\cite{strongmono} as it embodies the first
proof of strong monogamy beyond strong symmetry requirements; while
on the other hand, in the scope of the present paper, it enables us
to regard the considerations of the previous subsection as
mathematically valid over all the parameter space, and not only in
the limit of high squeezing $a$.

\section{Discussion: Where does promiscuity come from?}

\subsection{Some operational considerations}
From a practical point of view, two-mode squeezing transformations
are basic tools in the domain of quantum optics \cite{qopt} (they
occur e.g.~in parametric down-conversions), and the amount of
producible squeezing in  experiments is constantly improving
\cite{furuapl}. Only technological, no {\em a priori} limitations
need to be overcome to increase $a$ and/or $s$ to the point of
engineering excellent approximations to the demonstrated
promiscuous entanglement structure in multimode states of light
and atoms (see also \cite{suexp}). To make an explicit example,
already with realistic squeezing degrees like $s=1$ and $a=1.5$
(corresponding to $\sim$ 3 dB and 10 dB, respectively), one has a
bipartite entanglement of
$\tau(\gr\gamma_{1|2})=\tau(\gr\gamma_{3|4})=9$ ebits
(corresponding to a Gaussian entanglement of formation \cite{geof}
of $\sim 3.3\ $ebits), coexisting with a residual multipartite
entanglement of $\tau^{res}(\gr\gamma)\simeq 5.5$ ebits, of which
the tripartite portion is at most $\tau^{bound}(\gr\gamma _{1\vert
2\vert 3}) \simeq 0.45$ ebits. This means that one can
simultaneously extract at least $3$ qubit singlets from each pair
of modes $\{1,2\}$ and $\{3,4\}$, and more than a single copy of
genuinely four-qubit entangled states (like cluster states).
Albeit with imperfect efficiency, this entanglement transfer can
be realized by means of Jaynes-Cummings interactions \cite{pater},
representing a key step for a reliable physical interface between
fields and qubits in a distributed QI processing network.

\subsection{Emergence of promiscuity from discrete to continuous variable systems, and beyond Gaussian states}

 By constructing a simple and feasible
example we have shown that, when the quantum correlations arise
among degrees of freedom spanning an infinite-dimensional space of
states (characterized by unbounded mean energy), an accordingly
infinite freedom is tolerated for QI to be doled out.   The
construction presented here can be straightforwardly extended to
investigate the increasingly richer structure of entanglement
sharing in $N$-mode ($N$ even) Gaussian states via additional pairs
of two-mode squeezing operations which further redistribute
entanglement. Notice once more that the promiscuous entanglement
distribution happens with no violation of the {\it a priori}
monogamy constraints that retain their general validity in quantum
mechanics. The motivation is the following. In the CV scenario,
naively speaking, if one writes from \ineq{ckwine} that the
entanglement between one mode and the rest is equal to the sum of
bipartite entanglements plus the residual multipartite entanglement,
one can in principle approach the limiting `fully promiscuous' case
``$\infty = \infty + \infty$''. The only restriction imposed by
monogamy is to bound the divergence rates of the individual
entanglement contributions as the squeezing parameters are
increased. Within the restricted Hilbert space of four or more
qubits, instead, an analogous entanglement structure between the
single qubits is strictly forbidden (one can either have ``$1=1+0$''
or ``$1=0+1$'').

Our results open original perspectives for the understanding and
characterization of entanglement in multiparticle systems. Gaussian
states with finite squeezing (finite mean energy) are somehow
analogous to discrete systems with an effective dimension related to
the squeezing degree. As the promiscuous entanglement sharing arises
in Gaussian states by asymptotically increasing the squeezing to
infinity, it is natural to expect that dimension-dependent families
of states will exhibit an entanglement structure gradually more
promiscuous with increasing Hilbert space dimension towards the CV
limit. Let us further clarify this point. In the bipartite,
pure-state instance, a two-mode squeezed Gaussian state
$\ket{\psi_{sq}(r)}$ with squeezing $r \equiv (\log d)/2$ has
exactly the same entanglement $E=\log d$ (quantified by the
logarithmic negativity \cite{vidwer}) as a maximally entangled state
of two qu$d$its, $\ket{\Phi_d} = d^{-1/2} \sum_{i=1}^{d} \ket{i i}
\in {\C}^d \otimes {\C}^d$. Thus, for any finite squeezing, Gaussian
entanglement is not, in a sense, taking advantage of the full
infinite-dimensional Hilbert space, but only of a restricted, finite
section of it. Only for $r \rightarrow \infty$, and accordingly $d
\rightarrow \infty$, the states $\ket{\psi_{sq}(r\rightarrow
\infty)}$ and $\ket{\Phi_{d \rightarrow \infty}}$ tend towards the
(unphysical) EPR state, and the entanglement and the mean energy
asymptotically diverge \cite{noteunlim}.

\begin{figure}[t!]
\includegraphics[width=6cm]{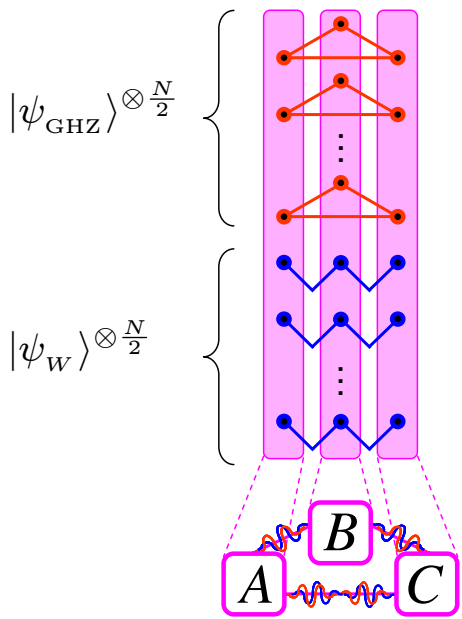} \caption{(Color online)
A tripartite pure state of three qu$d$its ($d=2N$ with $N$ even)
$A$, $B$, and $C$, is constructed as the tensor product of $N/2$
copies of a three-qubit GHZ state, and $N/2$ copies of a three-qubit
$W$ state. In the limit $d\rightarrow\infty$, $ABC$ constitute a
tripartite CV system whose (non-Gaussian) state exhibits full
promiscuity, i.e. unlimited genuine tripartite entanglement and
unlimited bipartite entanglement between any two parties.}
\label{figpromising}
\end{figure}

With such an analogy in mind, let us now construct the simplest
conceivable family of tripartite states, in which we will witness
promiscuity of distributed entanglement arising with increasing
dimensionality of the Hilbert space. We define the pure
permutationally-invariant state $\ket{\Psi_{2N}} \in {\C}^{6N}$ of a
system of $3N$ qubits (where $N \ge 2$ is an even integer, and we
label the state with the subscript ``$2N$'' for future convenience)
as the tensor product of $N/2$ copies of the three-qubit GHZ state
\cite{ghzs}, and of $N/2$ copies of the three-qubit $W$ state
\cite{wstates}. In formula,
\begin{equation}\label{psi3d}
\ket{\Psi_{2N}}=\ket{\psi_{_{\rm GHZ}}}^{\otimes \frac{N}{2}}
\otimes \ket{\psi_{_{W}}}^{\otimes \frac{N}{2}}\,,
\end{equation}
where $\ket{\psi_{_{\rm GHZ}}} =  \big(\ket{000} +
\ket{111}\big)/\sqrt2$ and $\ket{\psi_{_{W}}} = \big(\ket{001} +
\ket{010} + \ket{100}\big)/\sqrt3$. By grouping our $3N$ qubits into
three parties $A$, $B$ and $C$, each formed by $N$ qubits (each
qubit, in turn, taken from a single copy of either the GHZ or the
$W$ state), the state $\ket{\Psi_{d}} \in {\C}^{d} \otimes {\C}^{d}
\otimes {\C}^{d}$ of \eq{psi3d} becomes the tripartite state of the
three qu$d$its $ABC$ with $d \equiv 2N$. A graphical depiction of
such a construction is provided in Fig.~\ref{figpromising}. It is
straightforward now to evaluate the entanglement properties of
$\ket{\Psi_d}$. Using the tangle as an entanglement measure
\cite{notetangle} we find that the three qu$d$its $ABC$ share a
genuine tripartite residual entanglement ${\cal T}^{A|B|C}=d/4$, as
the GHZ state has unit three-tangle, and the $W$ state has zero
three-tangle \cite{ckw,wstates}. Similarly, the bipartite
entanglement between any two qu$d$its is ${\cal
T}^{A|B}_d=\tau^{A|C}_d=\tau^{B|C}_d=d/9$, as the two-qubit
reductions of $W$ states have a tangle of $4/9$, while the two-qubit
reductions of GHZ states are separable. Monogamy is clearly
satisfied (the tangle between one party and the other two is ${\cal
T}^{A|(BC)}=17d/36$), and one can see that both  -- tripartite and
every couplewise -- forms of entanglement are linearly increasing
with the dimension $d$, and are so monotonically increasing
functions of each other. Bipartite and genuine multipartite
entanglement can grow unboundedly (i.e., diverging for $d\rightarrow
\infty$), and are mutually enhanced: this is exactly what the
concept of promiscuity means.  Only for a three-qubit system
 there is mutual exclusion between maximum
tripartite entanglement (the GHZ case) and maximum two-party reduced
entanglement (the $W$ case). As soon as qu$d$its with $d>2$ are
concerned, an increasingly more freedom is let for entanglement
distribution compatibly with monogamy. An alternative demonstration
of the full promiscuity of entanglement in the state of \eq{psi3d}
is provided in Appendix \ref{appsq} by employing the `squashed
entanglement' \cite{Squashed} as bipartite entanglement monotone.

 From this simple example we
learn several things. First, that promiscuity gradually arises with
increasing dimension $d$ of the Hilbert space, mediating between the
prohibitive qubit scenario, and the libertine CV one. Second, that
the analogy between finite-squeezing CV states and
finite-dimensional systems can be extended to the multipartite
scenario. Third, and perhaps more interestingly, we see that
Gaussian states are too structurally simple as compared to general
non-Gaussian states of CV systems, and are hence sometimes unable to
achieve the full power of entanglement distribution in
infinite-dimensional systems even in the limit of a diverging mean
energy per mode. What we mean is quite straight. Taking the limit $d
\rightarrow \infty$ in the above qu$d$it picture, the state
$\ket{\Psi_{d\rightarrow \infty}}$ exhibits asymptotically diverging
entanglement both in genuine tripartite form and in couplewise form
between any two parties of the CV systems $A$, $B$ and $C$, i.e., an
{\em unlimited}, full promiscuous structure. We know instead that
three-mode Gaussian states can never achieve such a distribution of
entanglement \cite{contangle} (the reduced two-mode entanglement
stays finite even for infinite squeezing), and in this paper we had
to resort to four-mode states with broken symmetry to demonstrate
unlimited promiscuity in the Gaussian scenario.

The state $\ket{\Psi_{d\rightarrow \infty}}$ is indeed a highly
non-Gaussian state: a recently introduced quantifier  of
non-Gaussianity \cite{parisnongauss}, which measures the normalized
Hilbert-Schmidt distance $\delta$ between a non-Gaussian state and a
reference Gaussian state with the same first and second moments,
yields \cite{notenongauss}
\begin{equation}\label{parisianity}
\delta(\ket{\Psi_d}) = \frac{1}{2} + 2^{-\frac{3 d}{4} - 1} 3^{-d/4}
- 2^{d/2} 3^{-3 d/2} 7^{d/4}\,.
\end{equation}
On a scale ranging from zero to one, $\delta(\ket{\Psi_d}) \ge 0.48$
for any $d \ge 4$ (recall that $d=4$ corresponds to just one copy of
the GHZ state and one copy of the $W$ state), and increases with $d$
converging to $1/2$ for $d \rightarrow \infty$. As a term of
comparison, 1/2 is the asymptotic non-Gaussianity of the $n^{\rm
th}$ single-mode Fock state $\ket{n}$ for $n \gg 1$
\cite{parisnongauss}.  We have thus given the first evidence that
monogamy of CV entanglement holds beyond the Gaussian case (although
the state we considered is highly non generic, see the discussion in
\cite{lewebound}), and that, crucially, non-Gaussian states can
achieve a strictly more promiscuous structure in the sharing of
quantum correlations, than Gaussian states. More promiscuity means
more and stronger forms of entanglement encoded in different
multipartitions and simultaneously accessible for quantum
information purposes.

One can readily provide generalizations of the above example to
convince oneself that for an arbitrary number of qu$d$its in the
limit $d \rightarrow \infty$, a fully promiscuous structure of
entanglement is available (under and beyond permutation invariance)
for the corresponding multipartite non-Gaussian CV states.
Permutation-invariant Gaussian states (the so-called Gaussian
GHZ/$W$ states \cite{contangle}), instead, are always promiscuous
for any $N$, but never {\em fully} promiscuous, in the sense that
the reduced bipartite and/or $K$-partite entanglements ($K<N$) in
the various partitions always stay finite (and increasingly low with
higher $N$) even in the limit of infinite squeezing
\cite{strongmono}. And we stress that this is not a consequence of
restricting to a single mode per party: even considering a big $M
N$-mode CV system, partitioned into $N$ subsystems each comprising
$M$ modes, the corresponding $N$-partite Gaussian entanglement
(under permutation invariance) is exactly {\em scale invariant},
i.e., it is equal to the entanglement among $N$ single modes
(i.e.~to the case $M=1$) \cite{strongmono}, hence can never be fully
promiscuous. Of course, while Gaussian states cannot reach the
shareability of non-Gaussian states, mimicking the formers via the
latters is always possible. If one wishes to reproduce the $N$-mode
permutation-invariant Gaussian instance in terms of qu$d$its with
varying $d$, it is sufficient e.g.~for $N=3$ to modify the
construction of Fig.~\ref{figpromising} by unbalancing (as a
function of $d$) the fraction of GHZ versus $W$ copies, and keeping
the latter number finite while allowing the former to diverge with
$d$. In fact, as just shown, this Gaussian-like setting represents
clearly not the best entanglement sharing structure that states in
infinite-dimensional Hilbert spaces may achieve. It is worth
pointing out, however, that in the current experimental practice,
should a certain degree of promiscuity be required, it would be
surely easier to produce and manipulate Gaussian states of four
modes, with a quite high squeezing, than tensor products of many
three-qubit states.

\subsection{Concluding remarks and Outlook}

Inspired by the above discussion on the origin of entanglement
promiscuity, a more extended investigation into the huge moat of
qu$d$its appears as the next step to pursue, in order to aim at
developing the complete picture of entanglement sharing in many-body
systems \cite{pisa}. Here, we have initiated this program by
establishing a sharp discrepancy between the two {\it extrema} in
the ladder of Hilbert space dimensions: namely, entanglement of CV
systems in the limit of infinite mean energy has been proven
infinitely more shareable than that of individual qubits. We have
also given indication that such a transition between these two
diametrically opposite cases is not sharp, but smeared on the whole
scale of Hilbert space dimensionalities $2 < d < \infty$, as
promiscuity gradually and smoothly arises with increasing $d$.

Once a more comprehensive understanding will be available of the
distributed entanglement structure in high-dimensional and CV
systems (also beyond Gaussian states), the interesting task of
devising new protocols to translate such potential into full-power
QI processing implementations  can be addressed as well. In this
respect, it is worth remarking again that non-Gaussian states, which
by virtue of the extremality theorem \cite{extra} are systematically
more entangled that Gaussian ones at fixed second moments in a
bipartite scenario, appear also to have a richer structure when
entanglement sharing is addressed. In the tripartite case, for
instance, we have constructed an example of a fully promiscuous
non-Gaussian state, while promiscuity of three-mode Gaussian
entanglement is always partial. This provides a novel, additional
motivation for further characterizing non-Gaussian entanglement also
in the multipartite scenario, together with its possible
exploitation, e.g., to perform highly efficient teleportation
networks \cite{network} or to solve the Byzantine agreement problem
\cite{sanpera}.

\acknowledgements{We thank S. Bose, S. L. Braunstein, I.
Fuentes-Schuller, M. Paternostro, A. Serafini and P. van Loock for
discussions. Funding from the EU Integrated Project QAP IST-3-015848
is acknowledged. GA and FI are supported by CNR-INFM, INFN, MIUR; ME
is supported by The Leverhulme Trust.}

\appendix
\section{Non-Gaussian (full) versus Gaussian (partial) promiscuity of
tripartite entanglement sharing via the squashed
entanglement}\label{appsq}

Here we provide an alternative comparison (and contrast) between the
structure of entanglement sharing in the permutationally invariant
pure CV state obtained by taking the limit $d \rightarrow \infty$ of
the three-qu$d$it state in \eq{psi3d}, and the permutationally
invariant pure three-mode Gaussian states (Gaussian GHZ/$W$ states)
introduced in Ref.~\cite{contangle}. Even though more abstract, this
analysis is perhaps more conclusive as the same entanglement measure
is employed for both states. The {\em squashed entanglement}
$E_{sq}$ is an interesting information-theoretic entanglement
monotone which is, among the other properties, additive on tensor
product states \cite{Squashed}. Moreover, it has been proven to
fulfill monogamy, in the sense of \ineq{ckwine}, for arbitrary
$N$-partite systems, unregardingly of the associated Hilbert space
dimensions \cite{KoashiWinter}. Unfortunately, this measure is
uncomputable for almost all entangled states, even though it is
known to be upper bounded by the entanglement cost, and lower
bounded by the distillable entanglement (hence equal to the Von
Neumann entropy of entanglement on pure states). Let us bound the
different entanglement contributions in the state of \eq{psi3d} in
terms of squashed entanglement. The $1\times 2$ entanglement between
one qu$d$it, say A, and the other two, is
\begin{equation}\label{nongauss12squash}
E_{sq}^{A|(BC)}(\ket{\Psi_d})=(d/4)[E_{sq}^{A|(BC)}(\ket{\psi_{_{\rm
GHZ}}})+E_{sq}^{A|(BC)}(\ket{\psi_{_{W}}})] \approx 0.48 d
\end{equation}
(in this case the squashed entanglement reduces on each copy to the
computable Von Neumann entropy of one qubit). Concerning two-party
entanglement,  the GHZ states have all separable two-qubit
reductions; on the other hand, the bipartite squashed entanglement
in the reduced two-qubit states of each $W$ copy amounts to some
quantity, say $\omega$, which is strictly positive, as every
entangled two-qubit state has a nonzero distillable entanglement
(PPT criterion \cite{PPT}). The total bipartite squashed
entanglement between any two qu$d$its in the state \eq{psi3d} is
thus
\begin{equation}\label{nongauss11squash}
E_{sq}^{A|B}(\ket{\Psi_d}) = \omega d/4\,,\quad \omega > 0\,.
\end{equation}
From the general monogamy of the squashed entanglement, we
know that the residual three-partite entanglement is well defined
and positive,
$$E_{sq}^{A|B|C}(\ket{\Psi_d})=E_{sq}^{A|(BC)}(\ket{\Psi_d}) -
2E_{sq}^{A|B}(\ket{\Psi_d})\,.$$ Exploiting in particular the fact
that squashed entanglement is monogamous on each $W$ copy,
$E_{sq}^{A|(BC)}(\ket{\psi_{_{W}}}) \ge 2 \omega$, we immediately
obtain a lower bound on the three-qu$d$it residual entanglement,
\begin{equation}\label{nongauss111squash}
E_{sq}^{A|B|C}(\ket{\Psi_d}) \ge d/4\,,
\end{equation}
 where we have used the fact
that for each GHZ copy $E_{sq}^{A|B|C}(\ket{\psi_{_{\rm GHZ}}})=1$.
Looking at Eqs.~(\ref{nongauss12squash}--\ref{nongauss111squash}),
we observe once more that all forms of (squashed) entanglement are
mutually increasing functions of each other on the state of
\eq{psi3d}, and diverging in the limit $d \rightarrow \infty$. This
definitely proves that the corresponding asymptotic (non-Gaussian)
CV state, is definitely fully promiscuous, compatibly with the
monogamy of entanglement sharing.

On the other hand, let us repeat the same analysis for the Gaussian
  counterparts, represented by pure permutation-invariant
three-mode ``GHZ/$W$'' Gaussian states \cite{contangle} with
asymptotically diverging squeezing (and mean energy) in any single
mode. The $1\times 2$ squashed entanglement, equal to the Von
Neumann entropy of one mode, diverges with the squeezing (for $r \gg
0$ it goes $\sim 2r$). The reduced two-mode squashed entanglement,
instead, is smaller than the corresponding entanglement of formation
(which equates the entanglement cost as its additivity is proven in
this special case \cite{geof}). The latter increases with the
squeezing, converging to the finite value $\approx 0.278$ in the
limit $r \rightarrow \infty$. This unambiguosly prove that
permutation-invariant tripartite Gaussian states exhibit only a
partial promiscuity, as the genuine tripartite entanglement
(obtained by difference) diverges for infinite squeezing, while the
reduced bipartite entanglement stays constant. It is also known that
no three-mode Gaussian states can be more promiscuous than these
GHZ/$W$ states \cite{contangle}.


\begin{thebibliography}{99}

\bibitem{schr} E. Schr\"odinger, Proc. Cambridge Phil. Soc. {\bf 31}, 555 (1935).

\bibitem{book} {\em Fundamentals of
Quantum Information}, D. Heiss {\em ed.} (Springer, Berlin, 2002).

\bibitem{pisa} B. M. Terhal, IBM J. Res. \& Dev. {\bf 48}, 71
(2004);
G. Adesso and F. Illuminati,  Int. J. Quant. Inf {\bf 4}, 383
(2006).

\bibitem{ckw} V. Coffman, J. Kundu, and W. K. Wootters, Phys. Rev. A
{\bf 61}, 052306 (2000).

\bibitem{osborne} T. J. Osborne and F. Verstraete, Phys. Rev. Lett. {\bf 96},
220503 (2006).

\bibitem{contangle} G. Adesso and F. Illuminati, New J. Phys. {\bf 8}, 15 (2006).

\bibitem{hiroshima} T. Hiroshima, G. Adesso, and F. Illuminati,
Phys. Rev. Lett. {\bf 98}, 050503 (2007).


\bibitem{wstates} W. D\"ur, G. Vidal, and J. I. Cirac, Phys. Rev. A {\bf 62}, 062314 (2000).

\bibitem{strongmono} G. Adesso and F. Illuminati,
arXiv:quant-ph/0703277.


\bibitem{brareview} S. L. Braunstein and P. van Loock, Rev. Mod. Phys. {\bf 77}, 513 (2005).

\bibitem{menicucci} N. C. Menicucci {\it et al.}, Phys. Rev. Lett. {\bf 97}, 110501 (2006).

\bibitem{furuscience} A. Furusawa {\em et al.}, Science {\bf 282},
706 (1998).

\bibitem{slocc} M. Owari {\it et al.}, arXiv:quant-ph/0609167.


\bibitem{epr} A. Einstein, B. Podolsky, and N. Rosen, Phys. Rev. {\bf 47}, 777 (1935).

\bibitem{noteunlim} The notion of unlimited entanglement has to be
interpreted in the usual asymptotic sense. Namely, given an
arbitrarily large entanglement threshold, one can always pick a
state in the considered family with squeezing high enough so that
its entanglement exceeds the threshold.

\bibitem{qopt} S. M. Barnett and P. M. Radmore,
{\em Methods in Theoretical Quantum Optics} (Clarendon Press,
Oxford, 1997).

\bibitem{adebook} G. Adesso and F. Illuminati, arXiv:quant-ph/0701221, J. Phys. A (2007), in
press; J. Eisert and M. Plenio, Int. J. Quant. Inf. {\bf 1}, 479
(2003).


\bibitem{relaxandshare} J. Eisert, T. Tyc, T. Rudolph, and  B. C.
Sanders, arXiv:quant-ph/0703225.

\bibitem{ghzs} D. M. Greenberger, M. A. Horne, A. Shimony, and A. Zeilinger, Am.
J. Phys. {\bf 58}, 1131 (1990).



\bibitem{simon00} R. Simon, Phys. Rev. Lett. {\bf 84}, 2726 (2000).

\bibitem{werwolf} R. F. Werner and M. M. Wolf, Phys. Rev. Lett. {\bf 86}, 3658 (2001).

\bibitem{vidwer} G. Vidal and R. F. Werner, Phys. Rev. A {\bf 65}, 032314
(2002); J. Eisert, Ph.D. Thesis (University of Potsdam, 2001); M.
Plenio and S. Virmani, Quant. Inf. Comp. {\bf 7}, 1 (2007).


\bibitem{geof} M. M. Wolf {\it et al.}, Phys. Rev. A {\bf 69}, 052320 (2004).

\bibitem{ordering} G. Adesso and F. Illuminati, Phys. Rev. A {\bf 72}, 032334 (2005).

\bibitem{notefidel} The {\em
fidelity} ${\cal F} \equiv \bra{\psi^{in}}
\varrho^{out}\ket{\psi^{in}}$ (``in'' and ``out'' being input and
output state) quantifies the teleportation success \cite{brareview}.
For single-mode coherent input states and $\gr\gamma_{1|2}$ or
$\gr\gamma_{3|4}$ employed as entangled resources, ${\cal F}=
(1+e^{-2a}\cosh ^2s)^{-1}$.  It reaches unity for $a\gg 0$ even in
presence of high interpair entanglement ($s \gg 0$), provided that
$a$ approaches infinity much faster than $s$.

\bibitem{notemonocheck}
We need to check for monogamy explicitly because we are using the
contangle \cite{contangle} to quantify bipartite entanglement. Only
convincing numerical evidence, no fully analytical proof is yet
available for the monogamy of the contangle in nonsymmetric Gaussian
states of more than three modes. The general $N$-mode monogamy
inequality has been established instead for a related measure, known
as the Gaussian tangle \cite{hiroshima}. Monogamy proofs for the
contangle imply the corresponding proofs for the Gaussian tangle,
but the converse is not true.

\bibitem{furuapl} S. Suzuki {\it et al.}, Appl. Phys. Lett. {\bf 89}, 061116 (2006).

\bibitem{suexp} X. Su {\it et al.}, Phys. Rev. Lett. {\bf 98}, 070502 (2007).

\bibitem{pater} M. Paternostro {\it et al.}, Phys. Rev. A {\bf 70}, 022320
(2004).

\bibitem{notetangle}
The tangle ${\cal T}$ is an entanglement measure for qubit systems
\cite{ckw}. We need to extend its definition to arbitrary qu$d$its.
One important requirement is that the maximum entanglement should
scale with the dimension $d$, and not stay fixed to $1$; otherwise,
no fair comparison with the CV Gaussian states can be established.
The generalization of the tangle in terms of linear entropy
\cite{osborne} is thus not appropriate for this example. To our
aims, let us assiomatically define the `tangle' for qu$d$it systems
as the entanglement monotone which is (i) reduced to the usual
tangle on qubits and (ii) additive on tensor products states: ${\cal
T}(\rho^{\otimes n})=n {\cal T}(\rho)$. We do not need any further
specification for this measure, in order to evaluate it on the
example of Fig.~\ref{figpromising}. Notice, however, that the
physical conclusions we draw (in particular, the full promiscuity)
are not relying on the specific measure chosen to quantify
entanglement; see also Appendix \ref{appsq}.



\bibitem{Squashed}
M. Christandl and A. Winter, J. Math. Phys. {\bf 45},  829  (2004).




\bibitem{parisnongauss}
M. G. Genoni, M. G. A. Paris, and K. Banaszek, arXiv:0704.0639.

\bibitem{notenongauss}
It is straightforward to evaluate the non-Gaussianity for the state
in \eq{psi3d}. One has, for each single copy, $\delta(\psi_{_{\rm
GHZ}})= \frac{4769}{11664}  \approx 0.409$ and $\delta(\psi_{_{W}})=
\frac{5}{12}  \approx 0.417$, while the composition rule for
$\delta$ on tensor product states is given in
Ref.~\cite{parisnongauss}.

\bibitem{lewebound}  P. Horodecki and M. Lewenstein, Phys. Rev. Lett. {\bf 85},
2657 (2000).

\bibitem{extra}
M. M. Wolf, G. Giedke, and J. I. Cirac, Phys. Rev. Lett. {\bf 96},
080502 (2006).


\bibitem{network} P. van Loock and S. L. Braunstein, Phys. Rev. Lett.
{\bf 84}, 3482 (2000); G. Adesso and F. Illuminati, {\it ibid.} {\bf
95}, 150503 (2005); H. Yonezawa, T. Aoki, and A. Furusawa, Nature
{\bf 431}, 430 (2004).

\bibitem{sanpera} R. Neigovzen and A. Sanpera,
arXiv:quant-ph/0507249.

\bibitem{KoashiWinter}
M. Koashi and A. Winter, Phys. Rev. A {\bf 69}, 022309 (2004).

\bibitem{PPT} A. Peres, Phys. Rev. Lett. {\bf 77}, 1413 (1996);
M. Horodecki, P. Horodecki, and R. Horodecki, Phys. Lett. A {\bf
223}, 1 (1996).


\end{thebibliography}
\end{document}